\documentstyle[aps]{revtex}
\begin{document}
\draft
\title{Skyrmions in arbitrarily polarized quantum Hall states}
\author{Sudhansu S. Mandal$^1$ and V. Ravishankar$^2$}
\address{$^1$Department of Physics, Indian Institute of Science,
 Bangalore 560 012, India \\
 and Condensed Matter Theory Unit, Jawaharlal Nehru Centre for \\
 Advanced Scientific Research, Jakkur, Bangalore 560 064, India \\
 $^2$Department of Physics, Indian Institute of Technology,
 Kanpur 208 016, India}

\maketitle

\begin{abstract}

  We derive an effective non-linear sigma model for quantum hall
  systems with arbitrary polarizations, by employing the recently
  proposed doublet model. We study the topological excitations,
  in particular, the skyrmions, as a function of the filling fraction
  as well as the polarization. We determine the relationship between
  the topological charge density and the electronic charge density, and
  the statistics of skyrmions. We also estimate the value of spin 
  stiffness by using the dispersion relations that we have obtained
  recently by employing the time dependent Hartree-Fock approximation
  for the doublet model.
  Finally, we point out how the skyrmionic excitations reveal information
  directly on the  number of flux tubes that get attached to the electrons
  in order to form composite fermions.
\end{abstract}


Novel developments in semiconductor device fabrication have
led to interesting studies in multicomponent quantum Hall systems.
The components refer either to the spin degree in single layer systems
when the Zeeman energy is small (the so called vanishing Zeeman 
splitting limit),
or to the layer degree itself for multi layer systems. The layer degree
index can be effectively treated as a pseudo-spin index. Of course,
while the true spin couples to the magnetic field, the latter does not.
Nevertheless, there are many common features between the two systems
which allows a common treatment in many instances.

It has been seen experimentally \cite{expts} that the fully
polarized single layered quantum Hall states (QHS) at $\nu=1$ possess charged 
quasiparticle excitations with macroscopic size and large spin, 
verifying the predictions \cite{lee,sondhi,fertig}
that the lowest energy excitation for the state is the macroscopic
spin textured skyrmion. Subsequently, 
the presence of  similar topological excitations called `merons' has
been suggested \cite{macd} for describing a novel phase 
transition observed 
in double layer systems \cite{expt2}. In this paper, we propose to study
the former type of excitations for QHS with arbitrary polarization. We
necessarily consider single layered systems. We shall study the relation
between the topological and the electric charges, the charge and statistics of
the quasiparticles, estimate the spin stiffness, and finally the energetics.

Sondhi et al. \cite{sondhi} have described the skyrmionic excitations
in Laughlin $\nu =1/(2s+1)$ states by an effective nonlinear O(3) sigma
model, in terms of a unit vector field $\hat{n}(r)$ that represents the 
local spin polarization. 
They have shown that the topological charge density $q(r)$ of the skyrmions
is proportional to the particle density fluctuation and is given by
$\delta\rho (r)= \nu q(r)$, with $q(r)=\epsilon^{ij}\hat{n}\cdot
(\partial_i\hat{n}\times\partial_j\hat{n})/8\pi$. The topological
charge $Q=\int d^2r q(r)$ is an integer: positive for skyrmions and 
negative for antiskyrmions. The magnitude of $Q$ represents 
how many times the spin configuration wraps around a unit sphere, 
and its sign, the mode of wrapping.
For vanishing Zeeman energy, the skyrmions \cite{raja} are infinite
and their energy comes only through the spin stiffness. While the 
interaction between skyrmions favors macroscopic skyrmions, the
Zeeman term reduces their size.
The Laughlin states that Sondhi et. al. study are 
ferromagnetic in the vanishing Zeeman energy
limit, and the skyrmions are their lowest
energy charged excitations.
Recently Lilliehook et al. \cite{sondhi2} have reported that 
the skyrmions of charge $Q=2$ are in fact the lowest energy charged
excitations. Subsequently, Wu and Sondhi
\cite{wu} have found that the
skyrmions are not the lowest energy charged excitations in partially
polarized $\nu =3,5$ states, in agreement with Jain and Wu \cite{jain2}
who report that 
the fully polarized quasiparticles are the lowest energy excitations.
However, as we shall show below, the relation 
$\delta\rho (r) =\nu q(r)$ which
they have assumed in their analysis for partially polarized states is
not correct. Their above conclusion is correct though.

Starting from a Chern-Simon-Landau-Ginzburg (CSLG) Lagrangian for
composite bosons \cite{zhang},
Stone \cite{stone} has recently derived the nonlinear sigma model 
Lagrangian
which has been used by Sondhi et al \cite{sondhi} for Laughlin states.
Further Hansson et al. \cite{leinaas} have derived the sigma model 
Lagrangian for partially polarized states in a two component 
CSLG gauge theory 
of composite bosons. Their treatment is, however, not based on a
concrete model that captures all the known partially polarized states.
And the treatment of fully polarized and partially polarized
states is not unified.
In this paper, our goal is to derive an effective sigma model Lagrangian
for skyrmions on a single footing, irrespective of the polarization
of the ground state.
We accomplish this starting from 
a definite model, the doublet model \cite{model}, which has been
shown to correctly describe almost all known QHS with correct
polarization.

 To describe it briefly,
the model incorporates a doublet of U(1) Chern-Simons (CS) gauge
fields which couple to the fermions as follows:
\begin{eqnarray}
{\cal L} &=& \psi^\dagger_\sigma [i\partial_0 -ea_0^\sigma]\psi_\sigma
 -\frac{1}{2m}
\left\vert [\partial_j -ie(A_j+a_j^\sigma)]\psi \right\vert^2
+\frac{1}{2}\epsilon^{\mu\nu\lambda}\tilde{a}_\mu^\sigma 
\Theta^\prime_{\sigma\sigma^\prime}
\partial_\nu a_\lambda^{\sigma^\prime}
\nonumber \\
& & -\frac{1}{2}\int d{\bf \rm r}^\prime \delta \rho ({\bf \rm r})
 V(\vert {\bf \rm r}-{\bf \rm r^\prime }\vert )\delta \rho
({\bf \rm r^\prime}) +\frac{g}{2}\mu_B \psi^\dagger \vec{\sigma}\cdot \vec{B}
\psi  \; .
\label{eq0}
\end{eqnarray}
Note that in the CS term, apart from two fields which carry 
the spin index, we have a matrix valued strength
of the form $\Theta^\prime = \theta_1^\prime + \sigma_1 \theta_2^\prime$, 
where $\sigma_1$ is
the usual Pauli matrix.
In the diagonalized basis, one component of the CS gauge field
$a_\mu^+ =a_\mu^\uparrow + a_\mu^\downarrow$
interacts with fermions irrespective of spins while the other component
of the CS gauge field $a_\mu^- = a_\mu^\uparrow -a_\mu^\downarrow$
interacts in(out of) phase to the former with the
spin up(down) fermions. The respective strengths are given by
$\theta^\prime_+ =\theta^\prime_1 +\theta_2^\prime$ and
$\theta^\prime_- =\theta^\prime_1 -\theta_2^\prime$.
It is shown in Ref.[14] 
  that after rescaling of the CS 
gauge fields and their strengths,
if the strengths of the  
CS gauge fields $a_\mu^\pm$ are chosen to be $(e^2/2\pi)(1/2s)$ ($s$ is
an integer) and $\infty$ respectively by the 
requirement of composite fermions, the model describes arbitrarily
polarized QHS with correct spin polarization which is given by
\begin{equation}
\nu = \frac{p_\uparrow +p_\downarrow}{2s(p_\uparrow +p_\downarrow )+1}
\; ,\;\; \frac{\Delta \bar{\rho}}{\bar{\rho}}=\frac{p_\uparrow
-p_\downarrow }{p_\uparrow +p_\downarrow } \; .
\end{equation}
Here $p_\uparrow (p_\downarrow )$ is the effective number of Landau levels
filled by spin up(down) composite fermions, $2s$ is the even integral number
of flux quanta attached to each particle, and $\bar{\rho}$ and
$\Delta \bar{\rho}$ are the mean density and spin density respectively.
We have set the unit $\hbar =c =1$.
We, therefore, essentially have only one CS gauge field $a_\mu^+$ while
$a_\mu^-$ decouples.
We thus write the
Lagrangian density (dropping subscript and superscript
of the CS gauge field and its strength),
\begin{eqnarray}
{\cal L}[\psi ,a_\mu ] &=& \psi^\dagger [i\partial_0 -ea_0]\psi -\frac{1}{2m}
\left\vert [\partial_j -ie(A_j+a_j)]\psi \right\vert^2
+\frac{\theta}{2}\epsilon^{\mu\nu\lambda}a_\mu\partial_\nu a_\lambda
\nonumber \\
& & -\frac{1}{2}\int d{\bf \rm r}^\prime \delta \rho ({\bf \rm r})
 V(\vert {\bf \rm r}-{\bf \rm r^\prime }\vert )\delta \rho
({\bf \rm r^\prime}) +\frac{g}{2}\mu_B \psi^\dagger \vec{\sigma}\cdot \vec{B}
\psi  \; ,
\label{eq1}
\end{eqnarray}
where $\psi $ is a two-component component field of composite fermions
with effective mass $m$, $A_j$ is the vector potential for external magnetic
field. 
And $\delta\rho ({\bf \rm r})=\psi^\dagger ({\bf \rm r})\psi ({\bf \rm r})
-\bar{\rho}$ is density fluctuation over the mean density, $\bar{\rho}$,
of composite fermions interacting via the Coulomb potential $V(r)=e^2/
\epsilon r$. 

We now `transform' the composite fermion into a composite boson
by attaching
another unit of flux quantum $\phi_0= 2\pi /e$.
To accomplish this,
we introduce yet another  doublet of CS gauge fields, 
(in the same spirit of 
Ref.~\cite{model}
), with the coupling matrix
now given by
\begin{equation}
\Theta = \left( \begin{array}{cc}
	\theta_1 & \; \theta_2 \\
	\theta_2 & \; \theta_1 \end{array} \right) \; .
	\label{eq2}
\end{equation}
In the diagonalized basis, the Lagrangian density becomes
\begin{eqnarray}
{\cal L}[\psi , a_\mu ,a_\mu^+ ,a_\mu^-]
     &=& \psi^\dagger [ i\partial_0 -e(a_0+a_0^+ +\sigma_3 a_0^-) ]
   \psi \nonumber \\
   & & - \frac{1}{2m} \vert [ \partial_j -ie(A_j +a_j +a_j^+ +\sigma_3
   a_j^-)]\psi \vert^2   \nonumber \\
   & & +\frac{\theta}{2}\epsilon^{\mu\nu\lambda}a_\mu \partial_\nu a_\lambda
   + \frac{\theta_+}{2}\epsilon^{\mu\nu\lambda}a_\mu^+ \partial_\nu a_\lambda^+
   +\frac{\theta_-}{2}\epsilon^{\mu\nu\lambda}a_\mu^-\partial_\nu a_\lambda^-
   \nonumber \\
& & -\frac{1}{2}\int d{\bf \rm r}^\prime \delta \rho ({\bf \rm r})
 V(\vert {\bf \rm r}-{\bf \rm r^\prime }\vert )\delta \rho
({\bf \rm r^\prime}) +\frac{g}{2}\mu_B \psi^\dagger \vec{\sigma}\cdot \vec{B}
\psi  \; ,
\label{eq3}
   \end{eqnarray}
   where $a_\mu^\pm$ are the two extra CS gauge fields with respective
   coupling strengths $\theta_\pm = \theta_1 \pm \theta_2=
   (e^2/2\pi)(1/s_\pm)$, and $\psi $ is now a two-component boson field. 
   The requirement that the fermions transform into
  bosons imposes the constraint
  $s_+ + s_- =1$, which can be realized by two choices:
   (I) $s_\pm =1/2$,
   so that one flux quantum is seen by like spin bosons, and zero flux 
   is seen by unlike spin bosons,
   (II) $s_+ =1, \, s_- =0$ implying one unit of flux quantum is
   seen by all bosons, irrespective of
   spin. Other choices would lead to an anyonic 
   phase between particles of dissimilar spins. The second choice
   corresponds to decoupling of $a^-_\mu$ field. 
   We shall henceforth refer these choices by I and II respectively.

   In order to disentangle the charge and spin degrees of freedom, 
   we resolve $\psi $ as,
   \begin{equation}
   \psi = \sqrt{\rho} \phi \chi  
   \label{eq4}
   \end{equation}
   such that $\phi^\ast \phi =1 = \chi^\dagger \chi$. 
   Here $\chi$ is the CP$_1$ field which is related to the unit spin
   vector via $n^a = \chi^\dagger \sigma^a
   \chi$. By dropping the derivatives of the density $\rho$, we then obtain 
   \begin{eqnarray}
   {\cal L}[\phi , \chi , a_\mu , a_\mu^+ , a_\mu^-] 
   &=& \rho [i\phi^\ast \partial_0 \phi + i\chi^\dagger \partial_0 \chi 
   -e(a_0 + a_0^+ +n_3 a_0^- ]
   \nonumber \\
   & & -\frac{\rho}{2m}\vert i\phi^\ast \partial_k \phi +i\chi^\dagger
   \partial_k \chi -e(a_k+a_k^+ +n_3a_k^-)\vert^2  \nonumber \\
   & & +\frac{\theta}{2}\epsilon^{\mu\nu\lambda}a_\mu \partial_\nu a_\lambda
   + \frac{\theta_+}{2}\epsilon^{\mu\nu\lambda}a_\mu^+ \partial_\nu a_\lambda^+
   \nonumber \\
   & &+\frac{\theta_-}{2}\epsilon^{\mu\nu\lambda}a_\mu^- 
   \partial_\nu a_\lambda^-
   - \frac{\rho}{8m} (D_k^{ab}n_b)^2
   \nonumber \\
& & -\frac{1}{2}\int d{\bf \rm r}^\prime \delta \rho ({\bf \rm r})
 V(\vert {\bf \rm r}-{\bf \rm r^\prime }\vert )\delta \rho
({\bf \rm r^\prime}) +\frac{g}{2}\mu_B \rho n_3B \; ,
\label{eq5}
\end{eqnarray}
  where the covariant derivative is defined by
  $D_k^{ab}= \partial_k \delta^{ab} -2ie a_k^- L_z^{ab}$ with
  $L_z$ being the $z$-component of orbital angular momentum operator.
  The single derivative terms of $\phi$ and $\chi$ fields in Eq.~(\ref{eq5})
  can be described as gauge fields: e$A_{\mu\phi}=i\phi^\ast \partial_\mu
  \phi $ and e$A_{\mu \chi}=i\chi^\dagger \partial_\mu \chi $.

   Bose condensation of the composite bosons leads to quantum Hall state in
   the original electronic system \cite{zhang}. In other words, 
   the net mean field felt by the composite bosons is zero.
   Therefore the mean field configuration
   is taken to be
   $\langle B_\phi +B_\chi +B+b+b^+ +\tilde{b} \rangle = 0$,   
   where $\tilde{b} = \epsilon^{ij} \partial_i (n_3 a_j^-)$, and $B$'s and 
   $b$'s are the magnetic fields corresponding to respective gauge fields.
   The individual mean electric fields are also taken to be vanishing.

   In order to decouple the quartic term (second term in Eq.(\ref{eq5})),
   we introduce a Hubbard-Stratonovich field which is identified to be
   bosonic current $J_k(x)$. Therefore the effective Lagrangian for the
   fluctuating fields becomes
   \begin{eqnarray}
   {\cal L}[\phi ,\chi ,a_\mu , a_\mu^+ ,a_\mu^- ,J_k]
   &=& J_0 [i\phi^\ast \partial_0 \phi + i\chi^\dagger \partial_0
   \chi -e(a_0+a_0^+ +\bar{n}_3a_0^-)] \nonumber \\
   & & +J^k[ i\phi^\ast \partial_k \phi +i\chi^\dagger
   \partial_k \chi -e(a_k+a_k^+ +\bar{n}_3a_k^-)]  \nonumber \\
  & &  - \frac{1}{2K}\vert J^k \vert^2 -\frac{K}{8} (D_k^{ab}n_b)^2
    +\frac{\theta}{2}\epsilon^{\mu\nu\lambda}a_\mu \partial_\nu a_\lambda
   \nonumber \\
   & & + \frac{\theta_+}{2}
   \epsilon^{\mu\nu\lambda}a_\mu^+ \partial_\nu a_\lambda^+
  +\frac{\theta_-}{2}\epsilon^{\mu\nu\lambda}a_\mu^- \partial_\nu a_\lambda^-
  \nonumber \\
& & -\frac{1}{2}\int d{\bf \rm r}^\prime \delta \rho ({\bf \rm r})
 V(\vert {\bf \rm r}-{\bf \rm r^\prime }\vert )\delta \rho
({\bf \rm r^\prime}) +\frac{g}{2}\mu_B \bar{\rho} \bar{n}_3B \; ,
\label{eq6}
   \end{eqnarray}
   where $K=\bar{\rho} /m$, and the bosonic density $J_0 =\rho$.
   Now assuming that  $\phi$ is a pure phase given by 
   $\phi = e^{i\tilde{\theta}}$,
   we integrate over $\tilde{\theta}$ to obtain the continuity condition
   which can be realised as
   \begin{equation}
   eJ^\mu = \frac{1}{2\pi} \epsilon^{\mu\nu\lambda} \partial_\nu A_\lambda \; 
   , \label{eq7}
   \end{equation}
   where we have introduced an auxiliary field $A_\mu$.   
   Substituting this in the Lagrangian (\ref{eq6}), 
   we integrate over $A_0$ field in the
   gauge $\partial_j A_j =0$. 
   Subsequently we 
   perform the integrations (in Coulomb gauge)
   over $a_0$, $a_0^+$ and $a_0^-$, which
   respectively give the constraints 
   \begin{eqnarray}
   a_k &=& \frac{1}{2\pi}\frac{1}{\theta}A_k  \; ,\\
   a_k^+ &=& \frac{1}{2\pi}\frac{1}{\theta_+}A_k \; ,\\
   a_k^- &=& \frac{1}{2\pi}\frac{1}{\theta_-}\bar{n}_3A_k \; .
   \label{eq8}
   \end{eqnarray}
   Therefore the static Lagrangian becomes 
   \begin{eqnarray}
   {\cal L}[\chi ,A_k] &=& \rho (i\chi^\dagger \partial_0 \chi) -\frac{K}{8}
   (D_j^{ab} n_b)^2 +\frac{1}{2}g\mu_B
   \bar{\rho} \bar{n}_3 B \nonumber \\
   & & -\frac{K}{2}\int d^2 r \rho_{t}(r) \ln \vert r-r^\prime \vert
   \rho_{t}(r^\prime) \nonumber \\
& & -\frac{1}{2}\int d{\bf \rm r}^\prime \delta \rho ({\bf \rm r})
 V(\vert {\bf \rm r}-{\bf \rm r^\prime }\vert )\delta \rho
({\bf \rm r^\prime})  \; ,
\label{eq9}
   \end{eqnarray}
   where
   \begin{equation}
   \rho_{t}(r) = \frac{1}{2\pi e} \epsilon^{kl}\partial_l \left[ i\chi^\dagger
   \partial_k \chi -\frac{e}{2\pi}\left( \frac{1}{\theta}+\frac{1}{\theta_+}
   +\bar{n}_3\frac{1}{\theta_-}\right) A_k \right]
   \label{eq10}
   \end{equation}
   From the above two expressions (\ref{eq9} and \ref{eq10}),
   it is clear that for finite energy solutions, the topological
   charge density which has the form
   $q(r) \equiv \epsilon^{ij}\hat{n}\cdot (\partial_i \hat{n} \times
   \partial_j \hat{n} )/8 \pi $ asymptotically is
   \begin{eqnarray}
   q(r) &=& \frac{e^2}{2\pi}
   \left( \frac{1}{\theta}+\frac{1}{\theta_+} +\bar{n}_3
   \frac{1}{\theta_-}\right) \delta\rho (r) \nonumber \\
   &=& (2s+s_++\bar{n}_3s_-)\delta \rho (r)\; ,
   \label{eq11}
   \end{eqnarray}
   which displays the relation between topological charge density  
   and particle number density fluctuation.
   For smoothly varying configurations this relation will hold everywhere
   since the scale of variations is very large.
   Note that it depends both on the polarization 
   and the number of different kind of vortices attached to 
   each boson, in general. 
   For partially polarized QHS, the relation does not depend on the filling
   factor $\nu$, in disagreement with the assumption of Wu and
   Sondhi \cite{wu}. However, the relation is the same as $\delta \rho (r)
   =\nu q(r)$ for fully polarized $\nu =1/(2s+1)$ states as is obtained by 
   Sondhi et al. \cite{sondhi}
   The charge of the quasiparticles, i.e., skyrmions is then given by
   $e^\ast = e/(2s+s_+ +\bar{n}_3s_-)$, which we will come back to below.

   The total density of bosons is given by 
   $\rho (x) = \bar{\rho} +\delta \rho (x)$. Let us rewrite the term 
  $\delta \rho(x) (i\chi^\dagger \partial_0 \chi)$ in Eq.~(\ref{eq9})
  as
  \begin{equation}
  {\cal L}_H = \frac{e^{\ast 2}(2s+s_+ +\bar{n}_3s_-)}{2\pi} A_{0\chi}
  B_\chi \; .
  \label{eq12}
  \end{equation}
   ${\cal L}_H $ is identified with the Hopf term  which determines 
  the statistics of the skyrmions. 
  The statistics of the skyrmions is found to be $\pi /(2s+s_+ +\bar{n}_3
  s_-)$. In the case of partially polarized quantum Hall states, the
  statistical 
  phase depends on the polarization for the choice I; the statistics depends
  on the choice of the parameters $s_\pm $. However, for fully polarized
  Laughlin $1/(2s+1)$ states, either choice of $s_\pm $ gives the same
  statistics $\pi /(2s+1)$, in agreement with previous findings 
  \cite{yang,nayak} but in disagreement with Ref.~\cite{stat} where
  the statistics is claimed to be dependent on the spin of skyrmions.

  The final form of the Lagrangian is
   \begin{eqnarray}
   {\cal L}[\chi ,A_k] &=& {\cal L}_H + \bar{\rho}
      (i\chi^\dagger \partial_0 \chi) -\frac{K}{8}
   (D_j^{ab} n_b)^2 +\frac{g}{2}\mu_B\bar{\rho}\bar{n}_3B \nonumber \\
    & & -\frac{1}{2(2s+s_+ +\bar{n}_3s_-)^2}
    \int d^2r^\prime q(r) V(r-r^\prime) q(r^\prime) \; .
   \label{eq13}
   \end{eqnarray}

   Now, if we consider a general, static, rotationally
symmetric spin vortex solution \cite{leinaas}:
\begin{equation}
\chi (\vec{r}) = \left( \begin{array}{c} \cos \frac{\alpha (r)}{2}
 \\ \sin \frac{\alpha(r)}{2}e^{im\phi} \end{array} \right)
\,\, ,\,\, \vec{a}^- = a(r) \hat{e}_\phi \; ,
\label{eq14}
\end{equation}
then the unit spin vector
\begin{equation}
\hat{n} = [\sin \alpha (r) \cos (m\phi ),\, \sin \alpha (r)\sin
(m\phi ),\, \cos \alpha (r) ] \; .
\label{eq15}
\end{equation}
Given this configuration, the finite energy requirement implies that
asymptotically $(D_k^{ab} n_b)^2 =0$. Thus,
$a(r) \rightarrow (-m)/2r$ and  $ \cos [\alpha (r)] = \bar{n}_3$ as $r
\rightarrow \infty $. Note however that for the choice II, $a(r)$
does not exist at all.  
The resultant configurations 
of the skyrmions are the following.

(i) {\it Fully polarized states:} The physical space is compactified to
a sphere of unit radius such that $r=0$ corresponds to south pole
$(\alpha (0)=\pi )$ and $r=\infty$ becomes north pole $(\alpha (\infty)
=0)$. The spin at the north pole will be up while it is down at the south
pole. The unit sphere of spins wraps the physical
unit sphere exactly once, for skyrmions of topological charge $Q=\pm 1$.

(ii) {\it Partially polarized states:} The skyrmions are not the usual ones
in this case. Since $\cos \alpha (\infty) =\bar{n}_3 <1$, $r =\infty$
does not become north pole, but corresponds to an angle
$\cos^{-1}(\bar{n}_3)$. The unit sphere of spins wraps exactly once inside
the physical sphere for $Q\pm 1 $ in this case as well.

(iii) {\it Unpolarized states}: The skyrmionic excitations are not possible
in this case.

The second term in Eq.~(\ref{eq13}) can be expressed \cite{stone} as
$(\bar{\rho}/2)\vec{{\cal A}}\cdot \partial_0\hat{n}$ where $\vec{{\cal A}}
(\hat{n}) $ is the vector potential for a unit monopole. This term 
describes precession of spins in a tilted magnetic field. The effective
Lagrangian for the long wavelength static skyrmions now becomes
    \begin{eqnarray}
    {\cal L} [\hat{n}]
    &=& {\cal L}_H +\frac{\bar{\rho}}{2}\vec{{\cal A}}\cdot \partial_0 \hat{n}
   -\frac{\rho_s}{2}(\bigtriangledown \hat{n})^2 
   +\frac{g}{2}\mu_B\bar{\rho}\bar{n}_3 B \nonumber \\
    & & -\frac{1}{2(2s+s_+ +\bar{n}_3s_-)^2}
    \int d^2r^\prime q(r) V(r-r^\prime) q(r^\prime) \; ,
    \label{eq16}
    \end{eqnarray}
    $\rho_s$ is the spin stiffness which is normalized due to Coulomb 
    interaction. The second and third terms in (\ref{eq16}) is standard
    for a ferromagnetic system. The next two terms break the scale 
    invariance.
    The size and energy of the skyrmions depend crucially on the interplay
    between these two terms. As the value of the parameter
$\tilde{g} = g\mu_B B/(e^2/\epsilon l) \to 0$, the size of the skyrmions
increases while their energy decreases \cite{sondhi}. At $\tilde{g}=0$,
the skyrmions are of infinite extent having the energy $4\pi\rho_s $.

    The value of $\rho_s$ can be estimated by comparing dispersion relation
    from the above Lagrangian (\ref{eq16}) with the neutral long
    wavelength spin wave dispersion relation.
    The coefficient of the gradient term in the spin wave dispersion
    relation \cite{swave} is
     $k = \frac{1}{4} \sqrt{\frac{\pi}{2}} \frac{e^2}{\epsilon l_0}$ for
     fully polarized $\nu = 1/(2s+1)$ states and
     $k = \frac{7}{16} \sqrt{\frac{\pi}{2}} \frac{e^2}{\epsilon l_0}$ for
     partially polarized $3/(6s\pm 1)$ states. 
    Here $l_0$ is the effective magnetic length which is related with
    actual magnetic length via $(l_0/l)^2 = p/\nu$, where
    $p=p_\uparrow +p_\downarrow$ 
    is the effective number of Landau levels filled by CF.

    We now estimate spin stiffness $\rho_s = k/4\pi $, with the substitution
    $e \to e^\ast =e/(2s+s_++\bar{n}_3s_-)$. Therefore, for Laughlin states,
    \begin{equation}
    \rho_s = \frac{\sqrt{\nu}}{(2s+1)^2} \frac{1}{16\pi}\sqrt{\frac{\pi}{2}}
    \frac{e^2}{\epsilon l} \; .
    \label{eq17}
    \end{equation}
    Note that this result is obtained from the doublet model after incorporating
    the composite fermion requirement, and by 
    using the time dependent 
    Hartree-Fock approximation. Sondhi et al \cite{sondhi}
    estimated spin stiffness
    for $\nu =1/3$ and $1/5$ states using the results of single mode 
    approximation obtained by others \cite{sma}. They obtained 
    the values of $4\pi \rho_s =
    0.024$ and $0.006$ (in the unit of $e^2/\epsilon l$)
    for $\nu = 1/3$ and $1/5$ states respectively. These
    numbers are in good agreement with that of ours as $0.0201$ and 
    $0.0056$ respectively obtained from Eq.~(\ref{eq17}) here.

    For partially polarized states, the topological charge depends
    on the polarization as well for the choice I while it is independent
    of the polarization for the choice II. 
    Let us estimate the spin stiffness for $\nu =3/(6s\pm 1)$. It is given by
    \begin{equation}
    \rho_s = \left\{ \begin{array}{ll}
    {\Large
    \frac{\sqrt{\nu}}{\sqrt{3}}\frac{1}{(2s + 2/3)^2}
    \frac{7}{64\pi}\sqrt{\frac{\pi}{2}}\frac{e^2}{\epsilon l} }&
    {\Large \;{\rm for}\;s_+=\frac{1}{2},\;s_-=\frac{1}{2} }\\ 
    {\Large \frac{\sqrt{\nu}}{\sqrt{3}}\frac{1}{(2s + 1)^2}
    \frac{7}{64\pi}\sqrt{\frac{\pi}{2}}\frac{e^2}{\epsilon l} }&
    {\Large \;{\rm for}\;s_+=1,\;s_-=0 } 
    \label{eq18}
    \end{array}
     \right. \; .
\end{equation}

Let us now consider briefly the energetics of the excitations.
For fully polarized $\nu =1/(2s+1)$ states, the energy required to create a
fully polarized quasiparticle quasihole pair \cite{wu,swave}, $(\sqrt{\pi
\nu /2})(e^2/\epsilon l)$, is greater than that for creating a skyrmion
antiskyrmion pair of topological charge $Q=\pm 1$ at $\tilde{g} =0$, 
which is obtained from
Eq.~(\ref{eq17}) as $(\sqrt{\pi \nu /2})(1/2(2s+1)^2)(e^2/\epsilon l)$.
The ratio of the energies for these two kinds of charged excitations
is given by $E_{sk-ask}/E_{qp-qh} = 1/2(2s+1)^2 $.
Therefore, the skyrmions are the lowest energy charged excitations in
the ground state of fully polarized states.

For partially polarized $\nu =3/(6s\pm 1)$ states, 
the energy of fully polarized
quasihole quasiparticle pair \cite{swave}, $(3/4)(\sqrt{\pi \nu /6})
(e^2/\epsilon l)$ and the energy of skyrmion antiskyrmion pair,
$8\pi\rho_s$, which can be determined from Eq.~(\ref{eq18}). 
The ratio of these two energies is given by $E_{sk-ask}/E_{qp-qh}
=7/6(2s+2/3)^2$ and $7/6(2s+1)^2$ for choices I and II respectively.
Thus, the skyrmions are not the lowest energy charged excitations in partially
polarized integer states ($s=0$), 
in agreement with Wu and Sondhi \cite{wu}, although
there is a possibility of such excitations.
However, for both the choices, the skyrmions are, infact, the lowest energy
charged excitations in the partially polarized fractional states, 
contrary to the speculation of Wu and Sondhi \cite{wu}. 
We have pointed out that the prescription for obtaining 
composite bosons are not unique. The ambiguity has no impact on the
Laughlin states, but does matter for partially polarized states.
For these states,
the ambiguity of the
choices may be indeed resolved by experiments  by 
determining the excitation gap
for skyrmions in the limit 
$\tilde{g} \to 0$.

Finally, a brief remark on how a composite fermion parameter
gets determined by the study of skyrmionic excitations in
partially polarized phase.
We have recently shown \cite{test} that one of the composite fermion 
parameters, viz., $p$ - the number of Landau levels that are effectively
filled -
can be determined unambiguously by neutron scattering experiments in
unpolarized and partially polarized states. 
It follows from the above study that that the other
parameter, i.e, $2s$, the number of flux quanta attached 
to each composite fermion, gets determined independently,
by a determination of
the spin stiffness, in the vanishing Zeeman energy limit.
 Thus, we have, in principle, measurements that
can determine $\nu,p, s$ independently. If performed, they would provide a
decisive verification of the composite fermion model.

In summary, we have derived an effective Lagrangian which exhibits
skyrmionic excitations  in arbitrarily polarized quantum Hall states. 
We have determined
the relation between topological charge density and density fluctuation
and have found that their proportionality constant is not the filling factor
in general. The statistics of the skyrmions also is determined, which is
the same as that of Laughlin quasiparticles for fully polarized states. 
For partially polarized states, the statistics of skyrmions may depend
on polarization of the ground state as well.
Finally, we have estimated the values of spin stiffness for all the states,
and pointed out how the parameter $s$ gets determined independently.

{\it Acknowledgments:} SSM is grateful to Diptiman Sen for helpful discussions.
He also thanks JNCASR, Bangalore, for financial support.


\begin{references}

\bibitem{expts} S. E. Barrett, G. Dabbagh, L. N. Pfeiffer, K. W. West,
	       and R. Tycko, Phys. Rev. Lett. {\bf 72}, 1368 (1994);
	       {\it ibid.} {\bf 74}, 5112 (1995); A. Schmeller,
	       J. P. Eisenstein, L. N. Pfeiffer, and K. W. West, Phys. Rev.
	       Lett. {\bf 75}, 4290 (1995); E. H. Aifer, B. B.
	       Goldberg, and D. A. Broido, Phys. Rev. Lett. {\bf 76},
	       680 (1996); V. Bayot, E. Grivei, S. Melinte, M. B. Santos,
	       and M. Shayegan, Phys. Rev. Lett. {\bf 76}, 3479 (1996).
\bibitem{lee} D. H. Lee and C. L. Kane, Phys. Rev. Lett. {\bf 64},
	      1313 (1990).
\bibitem{sondhi} S. L. Sondhi, A. Karlhede, S. A. Kivelson, and E. H.
		 Rezayi, Phys. Rev. {\bf B47}, 16 419 (1993).
\bibitem{fertig} H. A. Fertig, L. Brey, R. Cote, and A. H. MacDonald,
		Phys. Rev. {\bf B 50}, 11 018(1994).
\bibitem{macd} K. Yang, K. Moon, L. Zheng, A. H. MacDonald, S. M. Girvin,
	      D. Yoshioka, and S.-C. Zhang, Phys. Rev. Lett. {\bf 72},
	      732 (1994); K. Moon, H. Mori, K. Yang, S. M. Girvin,
	      A. H. MacDonald, L. Zheng, D. Yoshioka, and S.-C. Zhang,
	      Phys. Rev. {\bf B 51}, 5138 (1995).
\bibitem{expt2} S. Q. Murphy, J. P. Eisenstein, G. S. Boebinger, L. N.
	      Pfeiffer, and K. W. West, Phys. Rev. Lett. {\bf 72},
	      728 (1994).
\bibitem{raja} See, R. Rajaraman, {\it Solitons and Instantons}, (North-Holland,
	     Amsterdam, 1982).
\bibitem{sondhi2} D. Lilliehook, K. Lejnell, A. Karlhede, and
      S. L. Sondhi, cond-mat/9704121
\bibitem{wu} X. G. Wu and S. L. Sondhi, Phys. Rev. {\bf B 51},
	     14 725 (1995).
\bibitem{jain2} J. K. Jain and X. G. Wu, Phys. Rev. {\bf B 49},
	5085 (1994).
\bibitem{zhang} S. C. Zhang, T. H. Hansson, and S. Kivelson,
	 Phys. Rev. Lett. {\bf 62}, 82 (1989).
\bibitem{stone} M. Stone, Phys. Rev. {\bf B 53}, 16 573 (1996).
\bibitem{leinaas} T. H. Hansson, A. Karlhede, and J. M. Leinaas,
	Phys. Rev. {\bf B 54}, R11 110 (1996).
\bibitem{model} S. S. Mandal and V. Ravishankar, Phys. Rev.
          {\bf B 54}, 8688 (1996); (E) {\it ibid.} {\bf B 56},
	   4284 (1997).
\bibitem{jain} J. K. Jain, Phys. Rev. Lett. {\bf 63}, 199 (1989).
\bibitem{yang} K. Yang and S. L. Sondhi, Phys. Rev. {\bf B 54},
	      R2331 (1996).
\bibitem{nayak} C. Nayak and F. Wilczek, Phys. Rev. Lett. {\bf 77},
	       4418 (1996).
\bibitem{stat} J. Dziarmaga, Phys. Rev. {\bf B 53}, 12 973 (1996).
\bibitem{swave} S. S. Mandal, Phys. Rev. {\bf B 56}, 7525 (1997).
\bibitem{sma} S. M. Girvin, A. H. MacDonald, and P. M. Platzman,
	     Phys. Rev. {\bf B 33}, 2481 (1986).
\bibitem{test} S. S. Mandal and V. Ravishankar, Phys. Rev. {\bf B 54},
               8699 (1996).

\end{references}
   \end{document}